\documentclass[
    ,final            
  ]
  {aipproc}

\layoutstyle{6x9}

\begin{document}

\title{3D non-LTE spectrum synthesis\\for Type Ia supernovae}


\author{M.~Kromer}{
  address={Max-Planck-Institut f\"ur Astrophysik, Karl-Schwarzschild-Str. 1,\\ D-85748 Garching b. M\"unchen, Germany}
}

\author{S.~A.~Sim}{
  address={Max-Planck-Institut f\"ur Astrophysik, Karl-Schwarzschild-Str. 1,\\ D-85748 Garching b. M\"unchen, Germany}
}

\author{W.~Hillebrandt}{
  address={Max-Planck-Institut f\"ur Astrophysik, Karl-Schwarzschild-Str. 1,\\ D-85748 Garching b. M\"unchen, Germany}
}

\begin{abstract}
Despite the importance of Type Ia supernovae as standard candles for cosmology and to the chemical evolution of the Universe, we still have no consistent picture of the nature of these events. Much progress has been made in the hydrodynamical explosion modelling of supernovae Ia in the last few years and fully 3-D explosion models are now available. However those simulations are not directly comparable to observations: to constrain explosion models, radiative transfer calculations must be carried out. We present a new 3-D Monte Carlo radiative transfer code which allows forward modelling of the spectral evolution of Type Ia supernovae from first principles, using hydrodynamical explosion models as input. Here, as a first application, we calculate line-of-sight dependent colour light curves for a toy model of an off-centre explosion.
\end{abstract}

\maketitle


\section{Introduction}
The greatest predictive power of explosion models for Type Ia supernovae can be extracted by 
performing parameter free radiative transfer calculations to directly link them to observational 
data. This requires a solution of the time-dependent 3-D radiative transfer problem in chemically
inhomogeneous models of supernova ejecta.

\section{Method}
We extended the 3-D Monte Carlo radiative transfer code introduced in 
\cite{Sim2007} to a non-grey opacity treatment following the ideas 
outlined in a series of papers by Lucy \cite{Lucy2002,Lucy2003,Lucy2005}. 
The basic assumption of this approach is that the ejecta are in homologous 
expansion thereby decoupling the radiative transfer from hydrodynamical 
explosion modelling. As input we therefore take densities, velocities 
and composition from explosion models which have been extended up to the 
phase of homologous expansion and map these on to a 3-D cartesian grid 
which expands with time. The total energy released by the synthesised 
$^{56}$Ni is divided into $N$ identical energy packets which are 
distributed on the grid according to the $^{56}$Ni distribution. These 
pellets follow the homologous expansion until they decay. Decay times are
sampled randomly according to the $^{56}$Ni$\rightarrow^{56}$Co$\rightarrow^{56}$Fe decay chain.\\
Upon decay, a pellet transforms to a $\gamma$-packet representing a bundle 
of monochromatic $\gamma$-radiation. The $\gamma$-packets then propagate
through the grid until they either interact with matter by Compton scattering, 
photoabsorption or pair creation or they leave the simulation volume. 
When a $\gamma$-packet interacts it can transfer its energy to an electron.
We assume that the timescale to thermalise these fast electrons is short and
convert the $\gamma$-packet into a packet of thermal kinetic energy, a 
$k$-packet in the nomenclature of Lucy \cite{Lucy2002,Lucy2003,Lucy2005}. 
$k$-packets are not propagated but are converted in situ to radiative energy
by sampling all the possible cooling channels, i.e. collisional excitation/ionisation, 
bound-free and free-free cooling.\\
In the case of bound-free or free-free cooling the $k$-packet is transformed
into a monochromatic energy packet representing ultraviolet-optical-infrared
(\textit{UVOIR}) radiation, a so-called $r$-packet. The frequency of the packet 
is sampled randomly in accordance with the selected cooling process. $r$-packets 
propagate through the grid either until they interact with matter by electron 
scattering, free-free absorption, bound-free absorption or line absorption or 
until they leave the simulation volume. For free-free absorptions the $r$-packets 
transform into $k$-packets, for line absorptions they transform into a packet of 
atomic internal energy while for bound-free absorption they may be converted to 
either type. Atomic internal energy is instantaneously transformed into 
thermal kinetic energy by collisional deexcitation/recombination or radiative
energy by line emission/radiative recombination by sampling the statistical 
equilibrium equations following the macro-atom approach of Lucy \cite{Lucy2002}.\\
When the simulation has finished, we extract the spectral evolution by binning 
the escaping $r$-packets by frequency, time and angle. Colour light curves 
are extracted from the spectral evolution by integrating the spectra over 
the filter functions.\\

In calculating opacities we need to know atomic level populations and therefore
the radiation field. In principle, these could be extracted exactly out of our 
simulation, but it is too computationally demanding to do so. Therefore, we
parameterise the radiation field in all grid cells in the nebular approximation
using radiation temperatures and dilution factors from Monte Carlo estimators 
(c.f. Mazzali \& Lucy \cite{Mazzali1993}). The ionisation balance is solved 
exactly using Monte Carlo estimators for the ionisation rates and the excitation 
state is computed assuming the Boltzmann distribution evaluated at a temperature
corresponding to the local energy density of the radiation field. Full details 
will be presented in a forthcoming paper.


\begin{figure}[p]
  \includegraphics{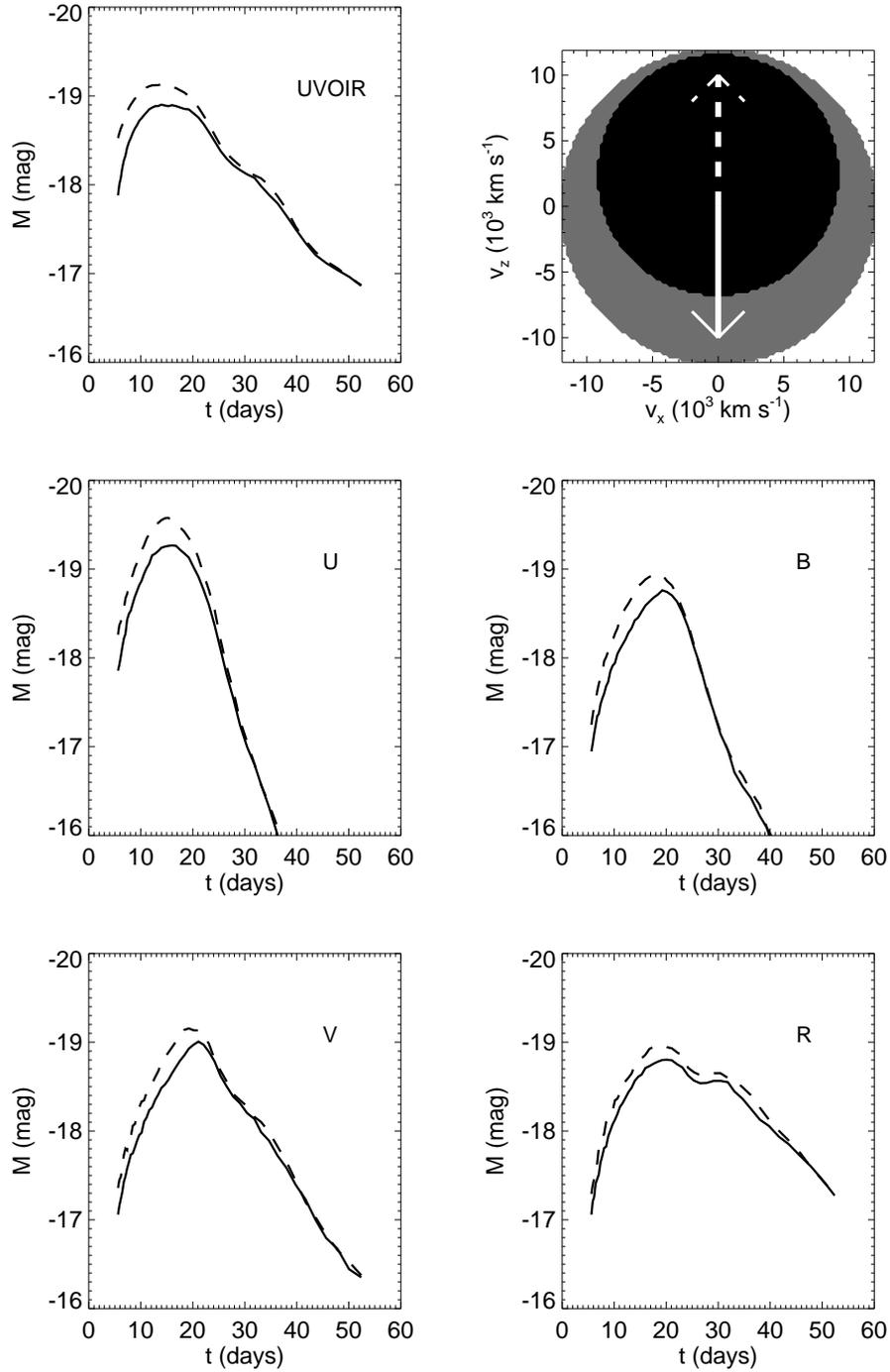}
  \caption{\textit{UVOIR} bolometric (top left panel), $U$ (middle left), $B$ (middle right), $V$ (bottom left) and $R$ (bottom right) light curves for the off-centre toy model. Dashed lines show light curves as seen from the Ni-rich side, solid lines those seen from the Ni-poor side. The top right panel shows the $^{56}$Ni distribution in the model's x-z plane. The black area contains $^{56}$Ni, the grey has no $^{56}$Ni. The arrows indicate the two lines-of-sight shown in the light curve plots.}
\end{figure}

\section{Application to an off-centre toy model}

After testing the code against other radiative transfer codes \cite{Kasen2006,Blinnikov1998} 
we have applied it to a parameterised off-centre explosion model (like those
in \cite{Sim2007a}) to make a preliminary study of line-of-sight effects on 
colour light curves. Our toy model has a total mass of 1.4~M$_\odot$ 
and uniform mass density. All the matter is confined in a ball with maximum
velocity $1.2\cdot10^4\;\mathrm{km\:s^{-1}}$. The $^{56}$Ni (0.52~M$_\odot$
in our model) is located in an inner ball. This Ni-rich ball is offset from
the centre of mass of the model by $2.4\cdot10^3\;\mathrm{km\:s^{-1}}$. Its
extension is chosen such that the $^{56}$Ni mass fraction inside the 
inner ball is 0.8.\\

We mapped this model onto a $50^3$ grid and followed the evolution of $10^7$
packets over 50 timesteps from 4 to 60 days after explosion. To reduce 
computational costs, we calculated the level populations in LTE and 
parameterised the radiation field by a black body adopting excitation and
radiation temperatures corresponding to the local energy density of the 
radiation field, which is extracted from a Monte Carlo estimator. We used 
atomic data from Kurucz CD 23 \cite{Kurucz1995}. Approximately $\sim3\cdot10^5$ 
atomic lines were used.\\

$U$, $B$, $V$, $R$ and \textit{UVOIR}-bolometric light curves for this model 
are shown in Figure 1 along the two most extreme lines-of-sight, i.e. from
the side in which the Ni bubble is displaced and from the opposite side. 
The light curves for intermediate lines-of-sight vary smoothly between those
extreme cases. As one expects the supernova appears brighter on the Ni-rich
side. The effect varies between 0.3 mag in $U$-band to 0.15 mag in $V$-band.
Furthermore light curves peak earlier on the Ni-rich side (see Table 1 for 
details). This confirms the trend seen in grey calculations (\cite{Sim2007a}
and Sim et al. in this volume) even though the effects are somewhat weaker
in our calculations. Further studies are necessary to probe a broader parameter
range and more realistic models as well as more complete atomic data sets.

\begin{table}
\begin{tabular}{llrrrr}
\hline
  &
  & \tablehead{1}{r}{b}{$U$}
  & \tablehead{1}{r}{b}{$B$}
  & \tablehead{1}{r}{b}{$V$}
  & \tablehead{1}{r}{b}{$R$}   \\
\hline
Ni-rich side: & Peak magnitude  & -19.58 & -18.94 & -19.16 & -18.97\\
 & Peak time (d)  &  15.30 &  19.24 &  19.24 &  19.24 \\
Ni-poor side: & Peak magnitude & -19.27 & -18.76 & -19.01 & -18.80\\
 & Peak time (d)  &  16.02 &  19.24 &  21.09 &  20.14 \\
\hline
\end{tabular}
\caption{Peak parameters of the colour light curves.}
\label{tab:a}
\end{table}

\section{Conclusion}
We presented a new time-dependent 3-D Monte Carlo radiative transfer code which
allows forward modelling of the spectral evolution of Type Ia supernovae from 
first principles, using hydrodynamical explosion models as input. As a first 
application we showed line-of-sight dependent effects on colour light curves 
for a toy model of an off-centre explosion. In future work we will extend this
study to more realistic models and the testing of state-of-the-art explosion
models.





\bibliographystyle{aipproc}   

\bibliography{kromer}

\IfFileExists{\jobname.bbl}{}
 {\typeout{}
  \typeout{******************************************}
  \typeout{** Please run "bibtex \jobname" to optain}
  \typeout{** the bibliography and then re-run LaTeX}
  \typeout{** twice to fix the references!}
  \typeout{******************************************}
  \typeout{}
 }

\end{document}